\documentclass[superscriptaddress,showpacs,pra,twocolumn]{revtex4}
\usepackage{epsfig}
\usepackage{delarray}
\usepackage{amsmath, amssymb}
\usepackage{bm}
\usepackage{dsfont}

\begin{document}
\title{Quantum Fisher Information of W States in Decoherence Channels}

\author{Fatih Ozaydin}
\email{MansurSah@gmail.com}
\affiliation{Department of Information Technologies, Isik University, Istanbul, Turkey}

\date{\today}
\begin{abstract}
We study the quantum Fisher information (QFI) of W states analytically with respect to SU(2) rotations in the basic decoherence channels i.e. depolarizing (DPC), amplitude damping (ADC) and phase damping (PDC), and present the interesting behavior of QFI of W states, especially when compared to that of GHZ states [Ma \textit{et al.,} Phys. Rev. A, \textbf{84,} 022302 (2011)]. 
We find that when initially pure W states are under decoherence, 
i) DPC: as decoherence starts and increases, QFI smoothly decays; 
ii) ADC: just as decoherence starts, QFI exhibits a sudden drop to the shot noise level and as decoherence increases, QFI continues to decrease to zero and then increases back to the shot noise level; 
iii) PDC: just as decoherence starts,  a sudden death of QFI occurs and QFI remains zero for any rate of decoherence, therefore W states in phase damping channel do not provide phase sensitivity.
We also find that, on the contrary to GHZ states, pure or decohered W states are not sensitive with respect to rotations in $z$ direction and the sensitivities with respect to rotations in $x$ and $y$ directions are equal to each other, implying no sudden change points of QFI due to competition between directions.

\end{abstract}
\maketitle


Quantum Fisher information (QFI) is the natural extension of Fisher information in the quantum regime and QFI of a parameter quantifies the sensitivity of a state with respect to changes of the parameter \cite{Braunstein1994PRL,Helstrom1976,Holevo1982}.
In particular, QFI characterizes the phase sensitivity of a state with respect to SU(2) rotations.
Therefore it plays a central role not only in estimation theory and but also in quantum information theory.
The limit on the variance of the estimation of a parameter $\phi$ of a general density matrix $\rho(\phi)$ is given by the quantum Cramer-Rao bound \cite{Helstrom1976,Holevo1982}

\begin{equation} 
\Delta \hat{\phi} \geq \Delta \phi_{QCB} \equiv  {1 \over \sqrt{ N_m F} }	 
\end{equation}
where $N_m$ is the number of experiments, $F$  is the quantum Fisher information and the estimator $\hat{\phi}$ satisfies $\langle \hat{\phi} \rangle = \phi$.
Considering $N_m = 1$, for separable states of $N$ particles, $F \leq N$, where equality holds for coherent spin states.
Therefore the precision limit of the estimation with the best separable states is $1 / \sqrt{N}$, which is called the shot-noise limit.
On the other hand, quantum Fisher information of a maximally entangled state, such as a pure GHZ state can reach $N^2$, implying the limit $1 / N$, which is the fundamental limit (also called the Heisenberg limit).
It was shown that QFI provides a sufficient condition to recognize multipartite entanglement:
If QFI of a state surpasses the shot-noise limit, then it is multipartite entangled and it is called a  ``useful'' state \cite{PezzeSmerzi2009PRL}. 
A basic property of multipartite entangled states is that they fall into inequivalent classes such as GHZ, W and Dicke states, and a state in one class cannot be converted to a state in another class via local operations and classical communication (LOCC) \cite{Dur2000PRA}, and for several tasks a specific multipartite entangled state is strictly required \cite{DHondt2006QIC}. Together with the discovery that not all multipartite entangled states exceeds the shot-noise limit -even when they are free of any decoherence \cite{Hyllus2010PRA,Hyllus2012PRA}, this property makes exploring QFI and the usefulness of each generic state a crucial step for quantum information science, especially when the state is subjected to decoherence due to natural effects. 
Recently, quantum metrology has been studied in non-markovian environments \cite{Plenio2012PRL} and in dissipative environments \cite{Berrada2012PLA,Rezakhani2014NJP,Alipour2014PRL}.
It was shown that the superpositions of pure Dicke states achieves larger QFI than pure Dicke states themselves  \cite{Xiong2010QIC}.  
We have studied the behavior of QFI of pure states in the superposition of GHZ and W states of several particles \cite{Ozaydin2013IJTP} and QFI of Bell states under decoherence \cite{Ozaydin2014APPA}.
QFI of NOON states in relativistic channels \cite{Kok2013PRA} and QFI of GHZ states in the basic decoherence channels  have been studied \cite{Wang2011PRA}. 
In the latter, it was found that in all three channels, there appears a competition between the phase sensitivities in each direction. Therefore QFI exhibits sudden change points. Also QFI of decohered GHZ states exhibit a smooth and continues decay starting from the QFI of the pure GHZ state.

In this work, we study the QFI of W states in the three basic decoherence channels, i.e. depolarizing, amplitude damping and phase damping. 
On the contrary to GHZ states \cite{Wang2011PRA}, we first show that, no matter being pure or decohered, W states do not provide phase sensitivity in $z$ direction and the phase sensitivities in $x$ and $y$ directions are equal to each other, which implies no sudden change points due to competition between directions. 
More interestingly we show that the phase sensitivity of W states under decoherence exhibit discontinuities such as sudden drop in amplitude damping channel and even sudden death in phase damping channel but exhibit a smooth and continues decay in depolarizing channel.

The maximal mean quantum Fisher information $\bar{F}_{max}$ of a possibly mixed state $\rho$ of $N$ qubits is given in \cite{Braunstein1994PRL,Wang2011PRA} as
\begin{equation} \bar{F}_{max}(\rho) = { c_{max} \over N} \end{equation}
where $c_{max}$ is the largest eigenvalue of the matrix \textbf{C} of which elements are given as
\begin{equation} \label{eq:Eq8}
   \textbf{C}_{kl} = \sum_{i \neq j} { (\lambda_i - \lambda_j)^2 \over \lambda_i + \lambda_j} [  \langle i| J_k |j \rangle \langle j| J_l |i\rangle + \langle i| J_l |j \rangle \langle j| J_k |i\rangle  ],
\end{equation}
where, $\lambda_{i,j}$ and ${|i\rangle,|j\rangle}$ are the eigenvalues and the associated eigenvectors of the density matrix of the state $\rho$ with the angular momentum operators on each particle in each direction, i.e.  
\begin{equation}
J_{\overrightarrow{n}} = \sum\limits_{\alpha=x, y, z} {1 \over 2} n_{\alpha} \sigma_{\alpha},
\end{equation} 
$\sigma_{\alpha}$ being the Pauli matrices.

In the case of $W$ states in these channels, matrix \textbf{C} appears as $diag\{ C_{xx}, C_{yy}, C_{zz} \} $ where $C_{xx}=C_{yy}$, $C_{zz}=0$ and the element $C_{kk}$ represents the phase sensitivity in $k$ direction.

Decoherence channels for a density matrix $\rho$ can be given in Kraus representation as \cite{NC}
\begin{equation}
\varepsilon ( \rho ) = \sum\limits_{\mu} E_{\mu} \rho E_{\mu}^{\dag}
\end{equation}
where the Kraus operators $E_{\mu}$ satisfy the completeness relation
\begin{equation}
\sum\limits_{\mu} E_{\mu}^{\dag} E_{\mu} = \mathds{1}.
\end{equation}
and $\mathds{1}$ is the $2x2$ identity matrix.

Below we will study QFI of $W$ states in the basic decoherence channels. For each decoherence channel, as in \cite{Wang2011PRA} we assume that each particle of the state is subjected to the same decoherence effect. We find that the general behavior of QFI of $W$ states does not depend on the number of particles $N$, and for the sake of simplicity we present the results for a $W$ state of three particles.

\begin{figure}[t]
  \centering
      \includegraphics[width=0.5\textwidth]{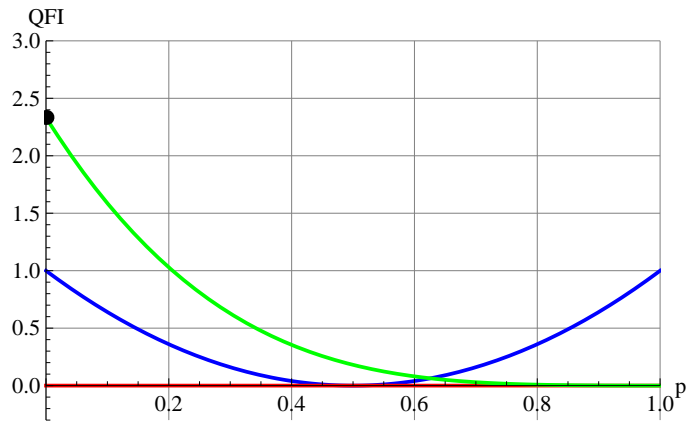}
  \caption{\label{fig:ColorHorizontalTree}(Color online). QFI of $|W_3\rangle$ in decoherence channels with respect to the decoherence strength $p$. Black dot is for a pure $|W_3\rangle$ state, i.e. $p=0$. Green, blue and red curves are for a $|W_3\rangle$ state in depolarizing, amplitude damping and phase damping channels, respectively.}
\end{figure}

\section{Depolarizing channel}
The Kraus operators of depolarizing channel for a single qubit are given by
\begin{equation}
\begin{split}E_0 = \sqrt{1- {3 \over 4} p} \mathds{1}, \ \  E_1 = \sqrt{{p \over 4}}\sigma_x, \\
E_2 = \sqrt{{p \over 4}}\sigma_y, \ \ E_3 = \sqrt{{p \over 4}}\sigma_z,
\end{split}
\end{equation}
and the eigenvalues of a $W$ state of three particles in the depolarizing channel appear as
$ \lambda_1 = {1 \over 8} (-2 + p)^2 p $;
$ \lambda_2 =  -{1 \over 8} (-2 + p) p^2 $;
$ \lambda_3 = \lambda_4 =    {1 \over 24} p (8 - 6 p + p^2)  $;
$ \lambda_5 =  {1 \over 24} p (16 - 24 p + 11 p^2) $;
$ \lambda_6 =   {1 \over 24} (24 - 52 p + 42 p^2 - 11 p^3) $ and
$ \lambda_7 = \lambda_8 =    {1 \over 24} (4 p - p^3)  $, with the associated normalized eigenvectors. We do not give the eigenvectors and the lengthy calculations but using Eq.(\ref{eq:Eq8}) it is straightforward to show that the maximal mean QFI of a $|W_3\rangle$ state in depolarizing channel, starting from the level of QFI of a pure $|W_3\rangle$ state, exhibits a smooth decrease with respect to the depolarization strength and vanishes when the depolarization strength is maximum (see the green curve in Fig.1). In the case of depolarizing channel, only the starting point and therefore the steepness of the decrease of the QFI of $W$ states depends on the number of particles, and this result is similar to that of GHZ states. 

\section{Amplitude damping channel}
The Kraus operators of the amplitude damping channel are given by
\begin{equation} 
E_0 = |0\rangle\langle0| + \sqrt{1-p} |1\rangle\langle1|, \ \ E_1 = \sqrt{p} |0\rangle\langle1| \end{equation}
where $p$ is the probability of decay from upper level $|1\rangle$ to the lower level $|0\rangle$ with the damping rate $\gamma$ i.e. $1-p = e^{-\gamma t /2 }$. We find the eigenvalues of the density matrix of a $W$ state of three qubits as: $\lambda_1 = 1-p$ and  $\lambda_2=p$, with the associated eigenvectors. Using Eq.(\ref{eq:Eq8}) we construct the \textbf{C} matrix and find the largest eigenvalue of $\textbf{C}$ matrix as $\lambda_{max} = 3 (1-2p)^2$, therefore the maximal mean quantum Fisher information of a $|W\rangle$ state in amplitude damping channel with decoherence strength $p$ as

\begin{equation}
  \bar{F}_{max}=\begin{cases}
    2.\overline{33}, & \text{$p=0$},\\
    (1-2p)^2, & \text{$0<p\leq 1$},
  \end{cases}
\end{equation}

which shows that, as plotted in Fig.1 (the blue curve), QFI of $W$ states exhibit a sudden drop to shot-noise level, when subjected to amplitude damping noise, and as the strength increases, QFI first vanishes and then increases back to shot-noise level.

\section{Phase damping channel}
The Kraus operators for the phase damping channel are given by

\begin{equation} E_0 = \sqrt{1-p} \mathds{1} , \ \ E_1 = \sqrt{p} |0\rangle\langle0|, \ \ E_2 = \sqrt{p} |1\rangle\langle1|. \end{equation}

In the phase damping channel, the eigenvalues of $|W_3\rangle$ state appears as
$ \lambda_1 = \lambda_2 = {1 \over 3} (2p -  p^2)$;
$ \lambda_3 = {1 \over 3} (3 - 4p  + 2 p^2)$ with the associated eigenvectors. Via Eq.(\ref{eq:Eq8}) we find that at any non-zero strength of phase damping, QFI vanishes, i.e.

\begin{equation}
  \bar{F}_{max}=\begin{cases}
    2.\overline{33}, & \text{$p=0$},\\
    0, & \text{$0<p\leq 1$},
  \end{cases}
\end{equation}

which implies a sudden death of quantum Fisher information, as plotted in Fig.1.

In conclusion, we have studied quantum Fisher information (QFI) of $W$ states with respect to SU(2) rotations under three decoherence channels and reported the interesting behavior of QFI of $W$ states when subjected to i) Depolarization: As decoherence starts and increases, QFI starts at the level of pure $W$ state, decreases smoothly and finally vanishes with full depolarization.  ii) Amplitude Damping: As the decoherence starts, QFI directly decreases to shot noise limit, and with the increasing decoherence, QFI first vanishes and then starts to increase, reaching the shot-noise level at full decoherence; iii) Phase Damping: At any rate of decoherence, QFI vanishes, implying a sudden death of QFI. We also found that on the contrary to GHZ states, $W$ states do not provide phase sensitivity in $z$ direction and the phase sensitivities in $x$ and $y$ directions are equal to each other. Therefore QFI of $W$ states do not exhibit sudden change points due to the competition between directions. Besides the decoherence effects, quantum Fisher information has also been studied considering photon losses \cite{Jin2013PRA,Tsang2013NJP}. On the other hand, an intense effort has been devoted on preparation of large-scale photonic W states in the ideal case where no practical imperfections are taken into account \cite{Ozdemir2011NJP,Bugu2013A,Yesilyurt2013A,Ozaydin2014arXiv}. Therefore we believe that our work may be useful for the efforts in preparing large scale $W$ states, as well as the quantum critical phenomena and percolation in quantum networks \cite{Acin2007,Kieling2007} when the unavoidable natural decoherence effects are taken into account. 


\end{document}